\documentclass[12pt,a4paper]{article}
\usepackage[latin1] {inputenc}
\usepackage[T1]{fontenc}
\usepackage[english]{babel}
\usepackage[dvips]{graphics}
\usepackage[dvips]{color}
\usepackage{relsize}
\usepackage{graphicx,amsmath,amssymb,amsfonts}
\title{Lifshitz formula for the Casimir force and the Gelfand-Yaglom theorem}
\author{C.~Ccapa Ttira$^{a}$,~
C.~D.~Fosco$^{b,c}$,~
and
F.~D.~ Mazzitelli$^{c,d}$\\
{\normalsize\it $^a$Instituto de Fisica - UFRJ}\\
{\normalsize\it CP 68528, 21941-972 Rio de Janeiro, RJ, Brazil}\\
{\normalsize\it $^b$Instituto Balseiro}\\
{\normalsize\it Universidad Nacional de Cuyo}\\
 {\normalsize\it R8402AGP Bariloche, Argentina.}\\
{\normalsize\it $^c$Centro At\'omico Bariloche}\\
{\normalsize\it Comisi\'on Nacional de Energ\'\i a At\'omica}\\
 {\normalsize\it R8402AGP Bariloche, Argentina}\\
{\normalsize\it $^d$Departamento de F\'\i sica}\\
{\normalsize\it Facultad de Ciencias Exactas y Naturales - Universidad de Buenos Aires}\\
 {\normalsize\it Ciudad Universitaria, Pabell\'on 1, 1428 Buenos Aires, Argentina.}}

\begin{document}
\date{}
\maketitle
\begin{abstract}
We provide a Quantum Field Theory derivation of Lifshitz formula for the
Casimir force due to a fluctuating real scalar field in $d+1$ dimensions.
The field is coupled to two imperfect, thick,  plane mirrors, which are modeled by 
background potentials localized on their positions.
The derivation proceeds from the calculation of the vacuum
energy in the Euclidean version of the system, reducing the problem to the
evaluation of a functional determinant.  The latter is written, via
Gelfand-Yaglom's formula, in terms of functions depending on the structure
of the potential describing each mirror; those functions encode the
properties which are relevant to the Casimir force and are the reflection 
coefficients evaluated at imaginary frequencies. 
\end{abstract}
\section{Introduction}\label{sec:intro}
Just a few years after Casimir's calculation of the attractive force between
ideal metallic plates \cite{Casimir}, Lifshitz derived his celebrated
formula for the interaction of bodies with arbitrary, frequency-dependent
dielectric functions \cite{Lifshitz}.  In the original calculation,
Lifshitz considered two disjoint media-filled half-spaces with plane and parallel
boundaries. The calculation was performed at finite temperature and the
final result for the interaction force was written in terms of the
dielectric functions that describe, macroscopically, the electromagnetic
properties of each media. 

The impressive refinement achieved in precision experiments measuring the
Casimir force stimulated a large amount of further theoretical
calculations. Indeed, to explain recent experimental results it has become
increasingly important to use models that describe the mirrors in a more
realistic way. For example, the dependence of the Casimir interaction on
the geometry, temperature, and macroscopic electromagnetic properties of
the interacting bodies is a subject of growing interest~\cite{reviews}.  

As an outcome of the sustained research effort on these topics, Lifshitz
formula has been generalized in several directions.  In this work, we shall
focus on its generalization to the case of stratified media, i.e. a set of
plane-parallel layers of different materials.  The interaction between the
different slabs can be computed and  expressed in terms of the
electromagnetic properties of the layers or, more specifically, in terms of
their reflection coefficients, which generally depend on the field's
frequency and momentum. 

Lifshitz formula, and some of its generalizations, have been derived using various
different theoretical approaches. In the original setting~\cite{Lifshitz},
Lifshitz introduced a random field in the Maxwell equations, in order to
describe the fluctuating sources in the materials. The interaction between
bodies was computed by evaluating the appropriate component of the stress
tensor, after assuming a local correlation for the random field.  
Besides, the existence of different media was taken into account by 
imposing the corresponding boundary conditions for the Green's functions 
of the electromagnetic field.  
This formula has been recently rederived~\cite{Philbin} using a canonical
quantization approach for macroscopic QED, in which the starting point is the action 
of the electromagnetic field, coupled to a set of reservoir oscillators. 
An alternative approach~\cite{Bordagetal}, is to quantize the electromagnetic field in the presence 
of the plane-parallel layers, computing the zero-point energy by following
Casimir's original line of thought. Then the boundary conditions on the modes of
the electromagnetic field determine the allowed eigenfrequencies, and the
summation of the zero-point energy of each mode is performed using the
argument theorem.  Note that, in this derivation, the validity of Lifshitz
formula in lossy media is not apparent. 

Lifshitz formula has also been derived in a quite different  context,
namely, considering the mirrors, and the spaces between them, as a quantum optical
network~\cite{genetetal}. Within this formalism it becomes possible to compute
the force between lossy mirrors,  and the final answer can be expressed
just in terms of the frequency-dependent reflection coefficients of each mirror. 
Remarkably, these coefficients are the only properties of the media that become
relevant to the computation of the vacuum force, in spite of the fact that
the same coefficient may be obtained from different kinds of mirrors.  

\noindent The Casimir force for absorbing media has also been considered in
Ref.~\cite{QOS},  where Lifshitz formula has been derived  using  the theory
of quantum open systems.

Another route to the computation of the Casimir force is to consider a
vacuum field in the presence of background potentials localized on the
mirrors~\cite{MIT,nos1}. These models can be justified from a microscopic
point of view taking into account the interaction of the internal degrees
of freedom of the mirrors with the vacuum field~\cite{nos1,nos2}. Using a
functional approach, the integration of the internal degrees of freedom of
the mirror produces an effective action for the vacuum field. That
effective action contains a potential that is different from zero just at
the positions of the slabs. These models may be used to reproduce
different boundary conditions for the fields which approach perfect ones
only under certain specific limits.
Although localized on each mirror, the potentials are in general nonlocal
in time as well as in the coordinates parallel to the mirror. In other
words, the boundary conditions for the Fourier modes of the field may depend 
on frequency or momentum.

In this
kind of approach, the zero-point energy is given by the functional determinant
of an operator containing the potentials that describe the media, and the
formal general expression seems to be strongly dependent on the specific shape of
the potentials involved. This is however, a puzzling circumstance.
As already mentioned,  Lifshitz formula is a quite general
expression for the Casimir force between two plane and parallel 
mirrors, and the force depends on the potentials only through their
reflection coefficients. In this paper we show, by providing a field theoretic
derivation of the Lifshitz formula,  that that is indeed the case.

\noindent The derivation relies upon the use of Gelfand-Yaglom formula for 
functional determinants~\cite{GY} (for a modern review, see \cite{reviewGY}).
As a by-product, we shall find an expression which also holds true
in situations where an uncritical application of Lifschitz formula may be
problematic, namely, potentials leading to bound states. 

This paper is organized as follows: in Section~\ref{sec:derivation} we
define the kind of system we consider and derive an expression for the
Casimir force. Then in Section~\ref{sec:discussion} we discuss some
consequences of the general result, considering some particular examples. Our conclusions are
presented in Section~\ref{sec:conc}.

\section{Derivation of the Casimir force}\label{sec:derivation}
Throughout this paper, we shall consider the case of a single, massive,
real scalar vacuum field, $\varphi$, in $d+1$ spacetime dimensions,
equipped with an Euclidean action $S(\varphi)$, which has the structure:
\begin{equation}\label{eq:defsphi}
{\mathcal S}(\varphi)\;=\; {\mathcal S}_0(\varphi)\,+\,{\mathcal
S}_I(\varphi)\;,
\end{equation}
where $S_0$ is the free field action
\begin{equation}
{\mathcal S}_0(\varphi)\;=\;\frac{1}{2} \int d^{d+1}x \big[ (\partial
\varphi)^2 + m^2 \varphi^2 \big] \;, 
\end{equation}
while $S_I$ describes the interaction between the field and two mirrors.
Those mirrors are assumed to be plane and parallel, with their normals
pointing along the $x_d$ direction (the remaining directions
will be denoted by $x_0$ and $x_\parallel =x_1,...x_{d-1}$).   Their interaction with $\varphi$ shall
be described here by a potential $V$, local in $x_d$,  which, in view of the previous
assumptions,will be a function of $x_d$ concentrated on the regions occupied
by the mirrors, and vanishing elsewhere. Thus the form
of $S_I$ shall be:
\begin{equation}\label{eq:defsi}
{\mathcal S}_I(\varphi)\;=\; \frac{1}{2} \, \int 
d^{d+1}x  \, d^{d+1}x'  \;  \delta(x_d-x'_d)
 V(x_d,x_0-x'_0,x_\parallel -x'_\parallel) \, \varphi(x)
 \,\varphi (x')\;,
\end{equation}
which, as well as $S_0$, is invariant under translations in all the
spacetime coordinates except $x_d$, namely, under the shifts $x_\mu \to x_\mu +
c_\mu$ with $\mu = 0, 1, \ldots, x_{d-1}$, $c_\mu = {\rm constant}$. Note that
we have included a non-local dependence of the potential on the temporal and 
parallel coordinates, in order to describe more general responses of the mirrors.

The fact that  $V$ is concentrated around each mirror, for $x_d \in
[a_1,b_1]$ and $x_d \in [a_2,b_2]$, say,  may be made explicit by writing
it in terms of two functions, $V_1$ and $V_2$, with support on the
intervals $[0,\delta_1]$ and $[0,\delta_2]$, respectively, such that: 
\begin{eqnarray}
V(x_d,x_0-x'_0,x_\parallel-x'_\parallel) \;&=&\; V_1(x_d-a_1) \lambda_1(x_0-x'_0,x_\parallel-x'_\parallel)\,
\nonumber\\ &+&\, V_2(x_d - a_2) \lambda_2(x_0-x'_0,x_\parallel-x'_\parallel)\;.
\label{V12}
\end{eqnarray}
The functions $\lambda_i$ characterize the response of each mirror. 
We regard the distances $\delta_i = b_i - a_i$,
\mbox{$i = 1, 2$} as the `sizes' (i.e., widths) of the mirrors, while 
$l \equiv a_2 - b_1$ is the distance between them. 

To give sense to the forthcoming steps, we confine the system to a $d$
dimensional spatial box (containing the mirror) such that the field
satisfies Dirichlet conditions on all the $2 d$ boundaries.  That box is assumed
to have sides of equal length, $L_\parallel$, for the $d-1$ coordinates
which are parallel to the plates: \mbox{$-\frac{L_\parallel}{2} \leq x_i
\leq \frac{L_\parallel}{2}$}, $i = 1,\ldots, d-1$, while for the remaining
spatial coordinate, $|x_d| \leq \frac{L}{2}$, with $L$ not necessarily
equal to $L_\parallel$. Besides, the $x_0$ coordinate is also assumed to
have a finite range, $x_0 \in [-T/2,T/2]$, and the field to satisfy
periodic boundary conditions on that interval. 

We have represented, in Figure 1, our previous conventions and notations about the
mirrors' configurations, from the point of view of the $x_d$ coordinate.
Regarding the shape of the potentials, the usual, potential barrier
case has been depicted in Figure 1, since that it the case when considering
imperfect Dirichlet conditions. However, the derivation below is
independent of that assumption, and indeed, one can even consider potential
wells. 
\begin{figure}[h!] 
\begin{center}
\begin{picture}(0,0)%
\includegraphics{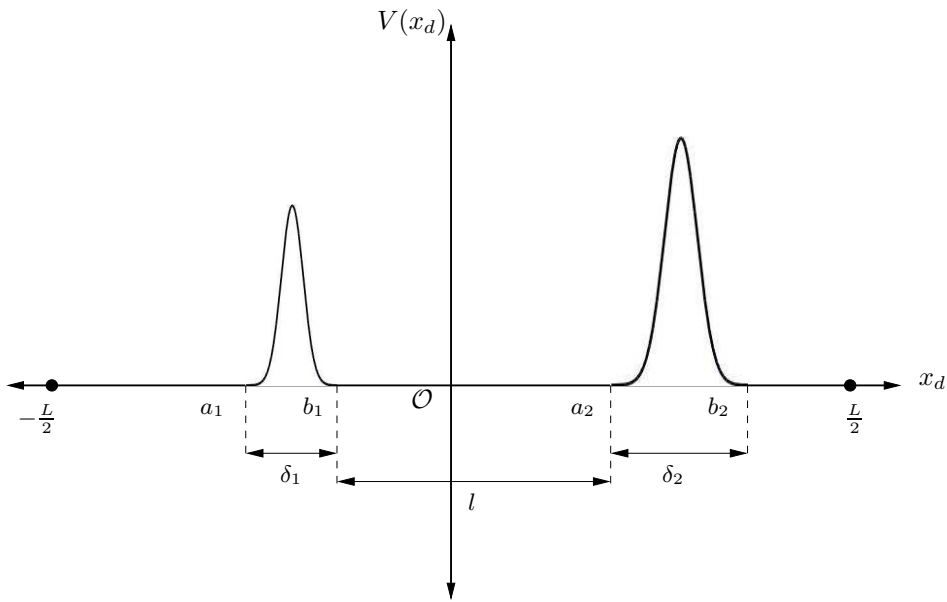}%
\end{picture}%
\setlength{\unitlength}{3108sp}%
\begingroup\makeatletter\ifx\SetFigFont\undefined%
\gdef\SetFigFont#1#2#3#4#5{%
  \reset@font\fontsize{#1}{#2pt}%
  \fontfamily{#3}\fontseries{#4}\fontshape{#5}%
  \selectfont}%
\fi\endgroup%
\begin{picture}(7237,4738)(1869,-4133)
\put(4816,434){\makebox(0,0)[lb]{\smash{{\SetFigFont{10}{12.0}{\rmdefault}{\mddefault}{\updefault}{\color[rgb]{0,0,0}$V(x_d)$}%
}}}}
\put(3421,-2626){\makebox(0,0)[lb]{\smash{{\SetFigFont{9}{10.8}{\rmdefault}{\mddefault}{\updefault}{\color[rgb]{0,0,0}$a_1$}%
}}}}
\put(4051,-3166){\makebox(0,0)[lb]{\smash{{\SetFigFont{9}{10.8}{\rmdefault}{\mddefault}{\updefault}{\color[rgb]{0,0,0}$\delta_1$}%
}}}}
\put(5086,-2581){\makebox(0,0)[lb]{\smash{{\SetFigFont{10}{12.0}{\rmdefault}{\mddefault}{\updefault}{\color[rgb]{0,0,0}${\mathcal O}$}%
}}}}
\put(7066,-3166){\makebox(0,0)[lb]{\smash{{\SetFigFont{9}{10.8}{\rmdefault}{\mddefault}{\updefault}{\color[rgb]{0,0,0}$\delta_2$}%
}}}}
\put(9091,-2401){\makebox(0,0)[lb]{\smash{{\SetFigFont{10}{12.0}{\rmdefault}{\mddefault}{\updefault}{\color[rgb]{0,0,0}$x_d$}%
}}}}
\put(4231,-2626){\makebox(0,0)[lb]{\smash{{\SetFigFont{9}{10.8}{\rmdefault}{\mddefault}{\updefault}{\color[rgb]{0,0,0}$b_1$}%
}}}}
\put(6346,-2626){\makebox(0,0)[lb]{\smash{{\SetFigFont{9}{10.8}{\rmdefault}{\mddefault}{\updefault}{\color[rgb]{0,0,0}$a_2$}%
}}}}
\put(7426,-2626){\makebox(0,0)[lb]{\smash{{\SetFigFont{9}{10.8}{\rmdefault}{\mddefault}{\updefault}{\color[rgb]{0,0,0}$b_2$}%
}}}}
\put(5536,-3391){\makebox(0,0)[lb]{\smash{{\SetFigFont{9}{10.8}{\rmdefault}{\mddefault}{\updefault}{\color[rgb]{0,0,0}$l$}%
}}}}
\put(1981,-2716){\makebox(0,0)[lb]{\smash{{\SetFigFont{9}{10.8}{\rmdefault}{\mddefault}{\updefault}{\color[rgb]{0,0,0}$-\frac{L}{2}$}%
}}}}
\put(8506,-2716){\makebox(0,0)[lb]{\smash{{\SetFigFont{9}{10.8}{\rmdefault}{\mddefault}{\updefault}{\color[rgb]{0,0,0}$\frac{L}{2}$}%
}}}}
\end{picture}%
\end{center}
\caption{The potential $V(x_d)$, using a typical profile for the mirrors.}
\end{figure}

The vacuum energy per $(d-1)$-dimensional mirror volume, ${\mathcal E}$,
relative to the vacuum energy in the absence of the mirrors ($V=0$), for a
finite $L$, can be expressed in terms of two Euclidean vacuum
persistence amplitudes:
\begin{equation}\label{eq:defe}
{\mathcal E} \;=\;\frac{1}{2} \, \lim_{T,L_\parallel \to \infty} 
\Big( \frac{1}{T L_\parallel^{d-1}} \, 
\log \frac{\mathcal Z}{{\mathcal Z}_0}\Big) \;,
\end{equation}
where ${\mathcal Z}$, the vacuum amplitude corresponding to $S$, can be written
as the functional integral:
\begin{equation}
 {\mathcal Z} \,=\, \int {\mathcal D}\varphi \; e^{- S(\varphi)} \;,
\end{equation}
while ${\mathcal Z}_0 \equiv {\mathcal Z}|_{V \to 0}$. Both path integrals
are performed over the space of fields vanishing on the boundaries of the spatial
box.

The exact -albeit formal- solution for the ${\mathcal Z}$ integral may then
be given in terms of a functional determinant:
\begin{equation}
{\mathcal Z} \,=\, \big(\det {\mathcal T} \big)^{-\frac{1}{2}} \;,
\end{equation}
with ${\mathcal T} \equiv  -\partial^2 + m^2 + V$, an operator acting on
functions of $x \in {\mathbb R}^{d+1}$ which vanish on all the boundaries of
the box, and $\partial^2 \equiv \partial_\mu \partial_\mu$, $\mu=0,1,\ldots,d$.

Separation of variables allows one to reduce the problem to another one for an
operator acting on functions of just one variable, $x_d$, but dependent on a
momentum vector $k=(k_0,\ldots,k_{d-1})$. The determinant becomes then block-diagonal:
\begin{equation}
\det {\mathcal T} \,=\,\prod_{k} \, \det \widetilde{{\mathcal T}}(k)
\end{equation}
with $\widetilde{{\mathcal T}}(k) = -\partial_d^2 + \Omega^2(k) + \widetilde {V}(x_d,k)$,
$\Omega(k) \equiv \sqrt{k^2 + m^2}$, which now acts on functions of $x_d$
which vanish at $x_d = \pm \frac{L}{2}$.  For a potential of the form
given in Eq.\ref{V12} we have
\begin{equation}
\widetilde{V}(x_d,k) \;=\; V_1(x_d-a_1) \widetilde{\lambda}_1(k)+ V_2(x_d - a_2) \widetilde{\lambda}_2(k)\equiv
\widetilde{V}_1(x_d,k) + \widetilde{V}_2(x_d,k)\;.
\label{Vtilde12}
\end{equation}

The values of $k$ become continuous and unbounded for the limit in~(\ref{eq:defe}), which we take
now: 
\begin{equation}\label{eq:defe1}
{\mathcal E} \;=\; \frac{1}{2} \, \int \frac{d^dk}{(2\pi)^d} \log 
\Big[\frac{\det \widetilde{\mathcal T}(k)}{\det \widetilde{\mathcal
T}_0(k)}\Big] 
\;,
\end{equation}
with $\widetilde{\mathcal T}_0 \equiv \widetilde{\mathcal T}|_{\widetilde{V}=0}$.  

It is interesting, at this stage, to note that the only change one should
make if one wanted to calculate the free energy per unit volume,
$\Gamma$, for the same system at a finite temperature $\beta^{-1}$ would be 
to keep the $x_0$ coordinate finite, with  $T = \beta$. The
integral over $k_0$ would then become a sum over the discrete Matsubara frequencies
$\omega_n = \frac{2\pi n}{\beta}$. Thus
\begin{equation}\label{eq:deff1}
\Gamma \;=\; \frac{1}{2\beta} \sum_{n=-\infty}^\infty
\int \frac{d^{d-1}k_\parallel}{(2\pi)^{d-1}} \log 
\Big[\frac{\det \widetilde{\mathcal T}(\omega_n,k_\parallel)}{\det \widetilde{\mathcal
T}_0(\omega_n,k_\parallel)}\Big] 
\;,
\end{equation}
where $k_\parallel=(k_1,\ldots,k_{d-1})$. Nevertheless, because the only
important difference between the evaluation of the energy or the free
energy emerges after the evaluation of the integrand, we just consider the
energy, commenting on the analogue result for the free energy at the end.

Note that the energy ${\mathcal E}$, besides being defined for a compact $x_d$
coordinate ($|x_d| \leq \frac{L}{2}$), still contains a contribution from 
the mirrors' self energies. One possible way
to get rid of those contributions is to consider the force (per unit
volume) ${\mathcal F}(L)$, and its $L \to \infty$ limit, the usual,
$L$-independent Casimir force, ${\mathcal F}_C \equiv {\mathcal F}(\infty)$:
\begin{equation}\label{eq:caf1}
{\mathcal F}(L)\;=\; - \frac{\partial{\mathcal E}}{\partial l} \;,\;\;\;
{\mathcal F}_C\;\equiv\; \lim_{L \to \infty} {\mathcal F}(L) \;.
\end{equation} 

To find the ratio between determinants appearing in (\ref{eq:defe}), we apply
Gelfand-Yaglom's (GY) theorem, which (after a rescaling and shift of
coordinates) allows we to write it as follows:
\begin{equation}
\frac{\det \widetilde{\mathcal T}(k)}{\det \widetilde{\mathcal T}_0(k)}
\;=\; \frac{\psi(\frac{L}{2})}{\psi_0(\frac{L}{2})} \;,
\end{equation}
where $\psi$ and $\psi_0$ are solutions, respectively, of the  homogeneous
equations:
\begin{equation}
\widetilde{\mathcal T}(k) \psi(x_d) \,=\, 0 \;\;, \;\;\;\;
\widetilde{\mathcal T}_0(k) \psi_0(x_d) \,=\, 0 \;, 
\end{equation} 
such that $\psi(-\frac{L}{2}) = 0$, $\psi'(-\frac{L}{2}) = 1$, and
identical conditions for $\psi_0$. 

Coming back to (\ref{eq:caf1}), we note that, 
\begin{equation}\label{eq:caf2}
{\mathcal F}(L)\;=\; - \frac{1}{2} \, \int \frac{d^dk}{(2\pi)^d} \,
\frac{\partial}{\partial l} \log
[\frac{\psi(\frac{L}{2})}{\psi_0(\frac{L}{2})}] \;,
\end{equation} 
the Casimir force obtained afterwards by taking the $L\to\infty$ limit.

In fact, because of the logarithmic derivative above, any $l$ dependent factor may be
dropped, in particular $\psi_0$, which is independent of $\widetilde{V}$, and hence of $l$. 

Thus the next step is the calculation of $\psi(\frac{L}{2})$ from the
second order differential equation; in what follows, we shall use $x$ to
denote the $x_d$ variable, since the problem is essentially
one-dimensional. 
We first note some properties of the solutions to the homogeneous
equation above, without making any assumption yet about $\widetilde{V}$, except that it
is regular enough as to make the existence and uniqueness theorem for the
second order homogeneous equation valid. 
Thus, given the values of $\psi$
and its derivative $\psi'$ at a single point, $x_i$, say, their values at any
other point, $x_f$, shall be uniquely determined. Introducing the
two-component $x$-dependent vector
$\Psi(x)=\left( \begin{array}{c} 
\psi(x) \\ \psi'(x)/\Omega \end{array}\right)$,
where we divided by $\Omega$ in order for the components to have equal
dimensions, after some (linear) algebra we may write:
\begin{equation}\label{eq:gsol1}
\Psi (x_f) \;=\; A(x_f,x_i) \, \Psi(x_i) 
\end{equation} 
where the matrix $A$ can be written as follows: 
\begin{equation}
A(x_f,x_i) \,=\, U(x_f) \, U^{-1}(x_i) 
\end{equation}
with 
\begin{equation}
 U(x) \;=\; \left( \begin{array}{cc}
\chi_1(x) & \chi_2(x) \\
\chi'_1(x)/\Omega & \chi'_2(x)/\Omega 
\end{array}\right) \;.
\end{equation}
Here, $\chi_1$ and $\chi_2$ denote two independent solutions of the
homogeneous equation, forming a  basis of solutions (it can be shown that
$A$ is independent of the choice of basis). From the fact that these two
functions are independent solutions of the homogeneous equation, it follows
that $\det U(x)$, their (scaled) Wronskian determinant, is a non vanishing constant. 
Therefore, we obtain the property:
\begin{equation}
\det A(x_f,x_i) \,=\, 1 \;.
\end{equation}

Note that $\psi(\frac{L}{2})$ may be written in terms of a single matrix elements of $A$, 
as follows:
\begin{equation}
\psi(\frac{L}{2}) \,=\, A_{12}(L/2, -L/2) \;,
\end{equation}
where we dropped $l$ independent factors.

For the particular case we are considering, namely, a potential which
vanishes everywhere except at the region occupied by the mirrors, the
previous expression may be rendered in a more explicit fashion. To that
order, it is convenient to make repeated use of (\ref{eq:gsol1}) to write
$A$ as a product of matrices corresponding to subintervals. Those factors
are completely determined by the potential inside the respective intervals;
thus, it is either just an $A_0$ factor, where the potential vanishes, or
it is determined by the solution of a homogeneous equation with $\widetilde{V}$
replaced by either $\widetilde{V}_1$ or $\widetilde{V}_2$. 
Thus:
$$A(L/2, -L/2) \;=\; A^{(0)}(L/2, b_2) \, A^{(V_2)}(b_2, a_2) $$
\begin{equation}\label{eq:gsol2}
\times A^{(0)}(a_2, b_1) \, A^{(V_1)}(b_1, a_1) \, A^{(0)}(a_1, -L/2) 
\end{equation} 
where we have indicated which potential each factor corresponds to. 
The matrix $A^{(0)}(x_f,x_i)$, corresponding to a null potential on the
interval $[x_i,x_f]$ can be immediately found, by using, for example, the
independent functions \mbox{$\chi_1(x) = e^{\Omega x}$} and \mbox{$\chi_2(x) =
e^{-\Omega x}$}, the result being: 
\begin{equation}
A^{(0)}(x_f,x_i)\,\equiv\,  A^{(0)}(\Delta x) \,=\, 
\left( 
\begin{array}{cc}
\cosh(\Omega \Delta x) & \sinh(\Omega \Delta x) \\
\sinh(\Omega \Delta x) & \cosh(\Omega \Delta x) 
\end{array}
\right) \;, 
\end{equation}
$\Delta x \equiv x_f - x_i$.
Besides, note that the only nontrivial factors are $A^{(V_{1,2})} \equiv
A^{(1,2)}$, $2 \times 2$ matrices depending on each potential, the wave
vector $k$; they depend on $a_i, b_i$ only through their differences,
$\delta_i$.

Taking into account the results above, we may write a quite compact
expression for $\psi(L/2)$:
\begin{equation}\label{eq:gsol3}
\psi(L/2) \;=\; u_2^T \, A^{(2)} \, A^{(0)}(l)\, A^{(1)}\, v_1 \;,
\end{equation} 
where we introduced the $2$-component vectors, $u_i$ $v_i$,
\begin{equation}
u_i \,=\, 
\left( 
\begin{array}{c} 
\cosh(\Omega l_i) \\
\sinh(\Omega l_i) 
\end{array}
\right)
\;,\;\;\;
v_i\,=\, 
\left( 
\begin{array}{c} 
\sinh(\Omega l_i) \\
\cosh(\Omega l_i) 
\end{array}\right) \;,
\end{equation}
and $l_1$, $l_2$ are the distances from the mirrors to the boundaries of the spatial box. 
We shall assume now that $b_1= -l/2$ and $a_2=l/2$, so that the internal
faces of the mirrors are symmetrically disposed with respect to the origin
(as well as the boundaries at $\pm L/2$). Then $l_{1,2} = (L-l)/2 -\delta_{1,2}$. 
Note that, in (\ref{eq:gsol3}), the only factors that depend on $l$ are
$u_2$, $v_1$, and $A^{(0)}$. 

To calculate the logarithmic derivative of $\psi(L/2)$, we first note that
\begin{eqnarray}
\frac{\partial \psi(L/2)}{\partial l}  &=& \Omega \Big[
 u_2^T A^{(2)} B^{(0)} A^{(1)} v_1 \nonumber\\
 &-& \frac{1}{2} \big(v_2^T  A^{(2)} A^{(0)} A^{(1)} v_1 
 + u_2^T  A^{(2)} A^{(0)} A^{(1)} u_1 \big) \Big]\;,
\end{eqnarray}
where
\begin{equation}
B^{(0)}\,\equiv\, \left( 
\begin{array}{cc}
\sinh(\Omega l) & \cosh(\Omega l) \\
\cosh(\Omega l) & \sinh(\Omega l) 
\end{array}
\right) \;. 
\end{equation}
Rather than writing the rather lengthy expression for the logarithmic
derivative for a finte $L$, we directly present its $\frac{L}{l} \to
\infty$ limit. Introducing the (constant) vector $w_1 = \frac{1}{\sqrt{2}} \left( \begin{array}{c} 1 \\
1 \end{array}\right)$, and matrix $C =  \left(\begin{array}{cc} -1 & 1\\ 1 & -1
\end{array}\right)$: 
\begin{equation}
\big[\frac{\partial \log \psi(L/2)}{\partial l}\big]_{L\to\infty}= \Omega e^{-
\Omega l} \frac{ w_1^T  A^{(2)} C A^{(1)} w_1}{w_1^T A^{(2)} A^{(0)} A^{(1)} w_1}
\;.
\end{equation}
It becomes clear from the last expression that it is convenient to perform
a change of basis in all the matrices involved. Indeed, rotating to the
basis $w_1$, $w_2$, with  $w_2 = \frac{1}{\sqrt{2}} \left( \begin{array}{c} 1 \\
-1 \end{array}\right)$, and denoting by $T^{(0,1,2)}$ and $D$ the form adopted by
$A^{(0,1,2)}$ and $C$ in the new basis, respectively, we see that:
\begin{equation}
\big[\frac{\partial \log \psi(L/2)}{\partial l}\big]_{L\to\infty}= \Omega e^{-
\Omega l} \frac{ w_1^T  T^{(2)} D T^{(1)} w_1}{w_1^T T^{(2)} T^{(0)} T^{(1)} w_1}
\;.
\end{equation}
More explicitly,
\begin{eqnarray}
\big[\frac{\partial \log \psi(L/2)}{\partial l}\big]_{\frac{L}{l}\to\infty} &=& 
\frac{-2 \Omega e^{- 2 \Omega l} \frac{T^{(2)}_{12}}{T^{(2)}_{11}}
\frac{T^{(1)}_{21}}{T^{(1)}_{11}}}{ 1
+ \frac{T^{(2)}_{12}}{T^{(2)}_{11}} \frac{T^{(1)}_{21}}{T^{(1)}_{11}} e^{- 2\Omega l}}
\nonumber\\
&=& \frac{\partial}{\partial l} \, \log \Big[ 1 +
\frac{T^{(2)}_{12}}{T^{(2)}_{11}} \frac{T^{(1)}_{21}}{T^{(1)}_{11}} 
e^{- 2\Omega l} \Big] 
\;.
\end{eqnarray}
Finally, we arrive to an expression for the force per unite volume,
\begin{equation}
{\mathcal F} \,=\, -\frac{1}{2} \int \frac{d^dk}{(2\pi)^d} \, 
\frac{\partial}{\partial l} \log \Big[ 1 +
\frac{T^{(2)}_{12}}{T^{(2)}_{11}} \frac{T^{(1)}_{21}}{T^{(1)}_{11}} 
e^{- 2\Omega l} \Big] 
\;,
\label{eq:final}
\end{equation}
which is the main result of this paper.

It is worth interpreting, at this point, the meaning of the ratios between matrix
elements of $T^{(1,2)}$ appearing in (\ref{eq:final}). From the original definition of
$A$, and implementing the change of basis, one sees that the matrix
elements of $T^{(i)}$, $i=1,2$  relate $\psi$ and $\psi'$ just to the right of
the mirror with the function and its derivative just to the left, but in a mixed
fashion. Indeed, in the new basis, it relates  functions such that 
$\psi' = \Omega \psi$ or  $\psi' = -\Omega \psi$ from one face of the
mirror to the other. Thus,  the matrix elements of $T^{(i)}$ connect
solutions of the form $e^{\pm \Omega x}$ on both sides of the mirror. 

We can make contact with the usual expression of Lifshitz formula from the
following observation: Identifying the Euclidean function $e^{ \Omega x}$
with the (Euclidean version of) a left moving mode, and $e^{- \Omega x}$
with a right moving mode, we see that
$r_R^{(1)}=\frac{T^{(1)}_{21}}{T^{(1)}_{11}}$ is the (Euclidean) reflection
coefficient associated to the left mirror, for incoming right modes. In the
same fashion,   $r_L^{(2)}=-\frac{T^{(2)}_{12}}{T^{(2)}_{11}}$ corresponds
to the reflection coefficient of the right mirror, for incoming right modes
(for 
non-symmetric potentials one may have $r_R^{(i)}\neq r_L^{(i)}$).   In terms of these coefficients
the Casimir force reads
\begin{equation}
{\mathcal F} \,=\, -\frac{1}{2} \int \frac{d^dk}{(2\pi)^d} \, 
\frac{\partial}{\partial l} \log \Big[ 1 -
r_R^{(1)}r_L^{(2)}
e^{- 2\Omega l} \Big] 
\; ,\end{equation}
which is of the same form that the Lifshitz formula for a scalar field.
As already mentioned in Section 2,  the case of finite temperature
is obtained by replacing
the integral over $k_0$ by a sum over dicrete Matsubara frequencies.

\section{Examples}\label{sec:discussion}
The matrices $T^{(i)}$ appearing in the final expression of the force can be explicitly 
computed in several particular cases. For instance, let us assume that a mirror is described
by a square potential barrier of height $\widetilde V_0(k)$ and width
$\delta$ \cite{piecewise}. A straightforward calculation yields:
\begin{eqnarray}
T_{11}&=& \cosh (\bar\Omega\delta)+\frac{\Omega^2+\bar\Omega^2}{2\Omega\bar\Omega}\sinh(\bar\Omega\delta)\nonumber \\
T_{22}&=&\cosh (\bar\Omega\delta)-\frac{\Omega^2+\bar\Omega^2}{2\Omega\bar\Omega}\sinh(\bar\Omega\delta)\nonumber \\
T_{12}&=&-T_{21}=\frac{\bar\Omega^2-\Omega^2}{2\Omega\bar\Omega}\sinh(\bar\Omega\delta)\, ,
\label{eq:piece}
\end{eqnarray}
where $\bar\Omega^2=\Omega^2+\widetilde V_0(k)$.  Note that for these symmetric potentials,
the left and right reflection coefficients coincide, that is $r^{(i)}_L=r^{(i)}_R$.

The Casimir force between two mirrors described by such potentials
is obtained by replacing the corresponding matrix elements into Eq.(\ref{eq:final}), and the final result is
in agreement with Ref.\cite{piecewise}. 
We show explicitly the expression of the force only in some limiting cases: for very thick slabs with $\delta_i >> l$, the vacuum
force reads
\begin{equation}
{\mathcal F} \,=\, -\frac{1}{2} \int \frac{d^dk}{(2\pi)^d} \, 
\frac{\partial}{\partial l} \log \Big[ 1 -
\frac{\widetilde V_0^{(1)} \widetilde V_0^{(2)}}{(\Omega+\bar\Omega^{(1)})^2(\Omega+\bar\Omega^{(2)})^2}
e^{- 2\Omega l} \Big] 
\; .\end{equation}
On the other hand, one can also obtain from the general expression (\ref{eq:piece}) the $T$ matrix
 associated
to a singular potential of the form $\widetilde V(x,k)=\widetilde \lambda(k)\delta(x)$, where
$\delta(x)$ is the Dirac $\delta$-function. Indeed, 
taking the limits $\widetilde V_0(k)\rightarrow\infty$ and $\delta\rightarrow 0$ with $\widetilde V_0(k)\delta\rightarrow
\widetilde\lambda(k)$,  the  $T$ matrix
becomes:
\begin{equation}
T =  \left( 
\begin{array}{cc}
1 + \frac{\widetilde\lambda}{2\Omega} & \frac{\widetilde\lambda}{2\Omega}  \\
- \frac{\widetilde\lambda}{2\Omega} & 1 - \frac{\widetilde\lambda}{2\Omega}
\end{array}
\right) \;, 
\end{equation}
thus, the force between two zero-width mirrors, with functions $\widetilde\lambda_1(k)$ and $
\widetilde\lambda_2(k)$, at a distance $l$ appart is
\begin{equation}
{\mathcal F} \,=\, -\frac{1}{2} \int \frac{d^dk}{(2\pi)^d} \, 
\frac{\partial}{\partial l} \log \Big[ 1 -
\frac{\frac{\widetilde\lambda_1\widetilde \lambda_2}{(2\Omega)^2}}{( 1 + \frac{\widetilde\lambda_1}{2\Omega})( 1 + \frac{\widetilde\lambda_2}{2\Omega})} 
e^{- 2\Omega l} \Big] 
\;,
\end{equation}
which reproduces a well known result \cite{nos2, piecewise,deltapot}.

\section{Conclusions}\label{sec:conc}

In this paper we presented a simple derivation of Lifshitz formula,
based on the use of GY theorem to compute the functional determinant
of an operator
of the form  $ -\partial^2 + m^2 + V$, where $V$ is a background potential
that models the imperfect mirrors.  

The fact that the Casimir force 
between flat mirrors
can be written just in terms of the reflection coefficients of each mirror, 
is quite transparent in this derivation.
Indeed, after 
separation of variables the problem reduces to the calculation of the
functional determinant of the one dimensional operator 
 $\widetilde{{\mathcal T}}(k) = -\partial_d^2 + \Omega^2(k) + \widetilde {V}(x_d,k)$.
GY formula implies that  this determinant is proportional to $\psi(L/2)$,
where $\psi(x_d)$ is a solution of  $\widetilde{{\mathcal T}}(k) \psi(x_d)=0$,
satisfying the initial conditions $\psi(-L/2)=0, \, \psi'(-L/2)=1$.
The key observation is that one can evaluate $\psi$
and its derivative on the right side of each mirror in terms of their values on the left
side, by means of a transfer matrix determined by the potential describing
the mirror.  
This matrix contains all the information of the potential upon which the
functional determinant may depend. On top of that, the Casimir force only depends on some 
particular ratio of elements of the transfer matrix.  Hence, the apparent
dependence on the function that describes the mirror collapses to a single
number (which may of course depend on frequency or momentum). 

We wish to point out that the self-energies of the mirrors do not appear
in the expression for the force, since the derivative with respect to the
distance between mirrors erases them out. The same would happen if we
considered the difference between  the energies corresponding to two
different distances.

Finally, we note that expression (\ref{eq:final}) for the force can be thought
of as a generalization of Lifshitz formula to the case of mirrors described by
potentials of the form $V= V(x_d,x_0-x'_0,x_\parallel-x'_\parallel)$, that
may include not only potential barriers but also potential wells. In this
case, equation (\ref{eq:final}) holds true, even when its interpretation in
terms of reflection coefficients in the real time formulation would be
problematic, because of the possible existence of bound states. Moreover,
potentials of this form are generated in a wide class of theoretical models in which 
the vacuum field is coupled to the internal degrees of freedom
of the mirrors, and therefore this approach provides an explicit  link between
the force and  the microscopic interactions.

\section*{Acknowledgements}
This work was supported by ANPCyT, CONICET, UBA and UNCuyo.

\end{document}